# An Accessible Toolkit for 360 VR Studies

Corrie Green*  
Robert Gordon University

Chloë Farr  
University of Victoria

Dr Yang Jiang  
Robert Gordon University

Dr John Isaacs  
Robert Gordon University

**ABSTRACT**

Virtual reality is expected to play a significant role in the transformation of education and psychological studies. The possibilities for its application as a visual research method can be enhanced as established frameworks and toolkits are made more available to users, not just developers, advocates, and technical academics, enhancing its controlled study impact. With an accessible first design approach, we can overcome accessibility constraints and tap into new research potential. The open-sourced toolkit demonstrates how game engine technologies can be utilized to immerse participants in a 360-video environment with curated text displayed at pre-set intervals. Allowing for researchers to guide participants through virtual experiences intuitively through a desktop application while the study unfolds in the user's VR headset.

**Index Terms**: I.3.8 [Computer graphics]: Application—Virtual Reality

## 1 INTRODUCTION

With the adoption of virtual reality becoming increasingly consumer focused, researchers must decide which VR headset meets their study requirements, such as tethered vs wireless and online vs offline support. Recent meta-reviews accessing more than 53 systematic reviews and meta-analyses support its use in anxiety disorder, pain management and weight related disorders showing its effect in clinical psychology [1]. Researchers designing these VR studies must initially decide what platform and software they will use. Some headsets require user logins and a requirement for both the participant and researcher to be in the same digital VR space. This proves difficult when documenting results, whilst also guiding the participant through the experience. With these considerations in mind of the researcher, how will the developed study be conducted at locations where headsets require an internet connection? What applications can be used?

AltspaceVR, Horizon Worlds, and VRChat are third-party software programmes that have been previously used to design VR studies [2]. These are social applications that can also be used to help researchers plan their studies utilizing environment design tools. However, the capability of these applications is limited. Affecting research design, participant satisfaction and result in unforeseen technical issues.

The developed toolkit could be a potential solution to this issue by providing an open source, intuitive and accessible interface to carry out visual studies. With a study in progress using its features for memory recall [3]. It is built using the Unity game engine and provides basic immersive interaction with 6DoF controllers and visualization of video and text. Technical knowledge of the game engine is therefore not required to use the compiled tool. Functionality *can* be addressed and added through the Unity Engine if desired, but for researchers who want to place a participant in an immersive environment with timed text and audio cues, the designed tool provides them with interactive guided sessions as such.

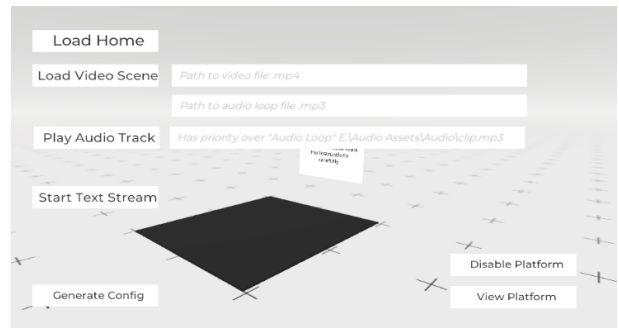

Figure 1: Main home interface displayed on the desktop to guide the session

## 2 ACCESSIBILITY

Commonly, crucial evaluations of interaction tactics are made at the discretion of the developer or researcher during the design of the environment, rather than through thorough end-user testing. The virtual reality (VR) sector still lacks a well-defined set of accessible design concepts to adhere to. A more efficient interaction system can be supported by a collaborative reduction of design options. Due to a reduction of features, first-time VR participants can have a reduced adjustment period without the pre-requisite of understanding controller features, or how to interact with specific components.

When using VR to conduct an immersive study, it is important to think about user accessibility. Headsets are traditionally heavy, wired, require physical movement to work with and are difficult for new users to operate. During a 360-degree experience, user head movement may be required while reading text. Focusing user attention in VR is problematic due to the 360 degree play space, thus a solution is to use graphical cues such as arrows visuals directing the user towards the area of interest such as subtitles. Similarly, without additional instruction, locating the source of in world audio can be difficult. Spatial sound will provide an omnidirectional auditory experience that will acoustically focus the user's attention to the desired direction.

However, persons hard of hearing may not be able to benefit from directional audio must also be considered. As a result, W3 [4] recommends that users have access to a mono audio sound option. Another method for directing user attention to the appropriate area is to utilise spatial arrows that point the user to where relevant information may be shown. Using the haptic rumbling in the motion controller to steer the participant to an area of interest is an alternate technique to draw attention to specific in-world areas of interest. These approaches have been considered in this solution.

## 2.1 Visual Elicitation

Having a controlled reproducible environment is clearly advantageous for psychological studies. VR visual-elicitation studies could help improve our understanding of environments including users' emotional responses to a presented situation. Visual elicitation is the method of conducting an interview with participants responding to visuals exhibited, which is traditionally done using a 2D image or video. They would describe the image's societal, personal, and value implications. Meanings conveyed by the participant from the images supplement the verbal discussion in such a way that without images then different emotions may be described [5].

## 3 TECHNICAL DISCUSSION

Used in this toolkit, the Unity XR plugin used to interface with the OpenXR framework is now becoming a widely adopted standard to support a one-build solution for compatibility with VR and AR headsets.

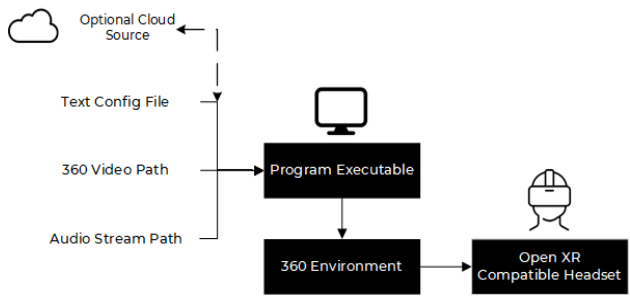

Figure 2: Data flow diagram from input files to VR headset

The architected solution supports streaming multiple audio files simultaneously, displaying text at timed intervals and allowing for 360-video playback. Due to the streaming mechanism using web requests, it is also possible to stream media to the environment with an internet connection. This option has been made available as 360-video files can have large file sizes. Building on this foundation of web requests, the application could be further enhanced by supporting more media types such as 3D model streaming.

By allowing the researcher to share the participant's perspective on a monitor, accurate guidance can be provided to the participant whilst also being able to update scenarios as desired. A text output is displayed front and centre to the play space's default position. A string and specified time interval can be sequentially read and displayed in VR from a configuration file. Another option is to play back words from a large body of text or book. It's been discussed if cognitive processes related to reading traditionally are complemented and reinforced by the tactile stimuli of the experience vs their digital counterpart [6]. This digital approach allows for audio, visual and haptic feedback if desired to compliment reading.

* https://orcid.org/0000-0003-0404-3668

## 3.1 Analytics

Metrics are not exclusively to be recorded by the researcher. Qualitative analysis of user movements can be recorded using the HMD orientation in relation to the text being displayed. This approach can also facilitate identifying areas of interest without the use of an eye tracker, but rather by approximating the user's gaze using the area they're facing in the 360-video. There is an opportunity to evaluate more than just the user's gaze to areas of the virtual environment. By fusing recorded user motion data with their biometric data such as the user's pulse, we can evaluate user's emotional response to a presented scenario with the capability to adjust the experience at runtime.

## 4 CONCLUSION

VR can support guided experiences for studies in healthcare to psychology with perception-based tasks. We searched for current design solutions that enable researchers to quickly develop text-based immersive studies and found limitations in guiding and directing a designed experience.

With increased research into accessible tooling, we can alleviate physical and cognitive accessibility issues for both researcher and participants, allowing for accurate, fresh research possibilities across scientific fields. Traditional photo-elicitation and walk-through interview procedures have failed to yield rich replies in previous photo-elicitation VR studies, however with further experiments and advances in accessible design this may no longer be the case [7]. By supporting best practices out of the box and focusing on accessible first design approaches, researchers from many fields can design and run a VR study effectively.

Download the project here

https://github.com/corriedotdev/vr-360-player